\documentclass[review]{elsarticle}

\usepackage{hyperref}

\journal{}

\begin{document}

\begin{frontmatter}

\title{Continuum Damage Model for Hydrogen Embrittlement in Ferritic Steels}

\author{Dakshina M. Valiveti\corref{mycorrespondingauthor}}
\cortext[mycorrespondingauthor]{Corresponding author}
\ead{dakshina.m.valiveti@exxonmobil.com}
\author{T. Neeraj}
\address{ExxonMobil Technology and Engineering, 1545 Route 22 East, Annandale, NJ 08801}

\begin{abstract}
Hydrogen embrittlement of metals and alloys, particularly steels, has been an important scientific and engineering challenge in the Oil and Gas industry for many years. It impacts the integrity and performance of a wide range of structures and equipment such as downhole tubulars and pipelines in sour service in the Upstream (U/S) and hydro-processing reactors in the Downstream (D/S). In addition, the rapidly growing interest in hydrogen as an energy carrier for fuel cells and mobility or as a clean fuel/heat source for hard to decarbonize industrial processes, draws attention to this key challenge of materials integrity in handling hydrogen. The fundamental understanding of failure mechanism(s) and the capability to model material behavior is important for cost effective materials selection, for managing the integrity and for repurposing existing infrastructure for transporting hydrogen as well as for extending the life of structures. To that extent, the present work develops a robust mathematical model to estimate the strength degradation and embrittlement due to hydrogen in steels. \\

The proposed model adopts the Nano-Void Coalescence (NVC) theory as the driving fracture mechanism for hydrogen embrittlement. NVC theory provides a more robust and comprehensive micro-mechanical pathway to failure than other hydrogen embrittlement theories in the literature(Neeraj {\it et. al,} Hydrogen embrittlement of ferritic steels: Observations on deformation microstructure, nanoscale dimples and failure by nanovoiding, Acta Materialia, 2012). A mathematical model is developed for hydrogen affected constitutive response of material, within the framework of finite element method. The modified constitutive response is a Gurson plasticity based continuum damage model and incorporates two vital aspects of NVC failure theory. These key aspects are (i) hydrogen enhanced localized dislocation plasticity and (ii) hydrogen enhanced vacancy stabilization forming nano-voids. The deformation and damage in the material is coupled with trap mediated hydrogen diffusion. The coupled model is implemented as a User Material (UMAT) in Abaqus finite element software. Validation and sensitivity analyses are performed on the developed model. Calibration of damage model parameters is performed for X65 commercial linepipe steel. Finally, capability of the damage model is demonstrated with numerical simulation of round bar tensile tests on X65 steel under hydrogen exposure. The numerical simulations are shown to be in excellent agreement with experimental results.
\end{abstract}


\end{frontmatter}

\section{Introduction}\label{intro}

The introduction of atomic hydrogen into metals and alloys during service is well known to degrade the mechanical properties in several applications in the form of hydrogen embrittlement. The sources for atomic hydrogen include (i) $H_2$ gas in pressure vessels and pipelines in hydrogen storage and transport applications for the hydrogen economy; (ii) high pressure $H_2$ gas in reactors; (iii) H generated from electrochemical reactions from an aqueous corrosion environment, particularly in sour (wet $H_2S$) service or from cathodic protection and (iv) H that is dissolved in the metal during material processing. In all of the above scenarios, hydrogen can reduce the ductility of the host metal/alloy significantly, depending on the amount of dissolved atomic hydrogen. There are multiple theories proposed in the literature attempting to explain the fundamental subcritical fracture mechanism(s) induced by hydrogen. While it is possible that a different mechanism is operative in different classes of material \cite{hirth,pouillier}, each of these theories has been developed based on observations on a subclass of alloy systems. Several review articles from recent literature \cite{rev-dwivedi,rev-dong, rev-laadel, rev-wu, rev-li, rev-meda} provide a very comprehensive overview of state of the art.
The proposed mechanisms for hydrogen embrittlement in steels can be broadly categorized into four main theories summarized below:

\begin{itemize}

\item{Hydrogen Enhanced Decohesion (HEDE): The decohesion theory is originally proposed by Troiano \cite{troiano} and later developed by Oriani \cite{oriani} , Gerberich and their coworkers. This theory proposed that when hydrogen accumulates locally within the lattice (for example due to a stress triaxiality), it reduces the cohesive Fe-Fe atomic bond strength. Therefore, it has been proposed that there exists a critical hydrogen concentration which when exceeded locally will lead to catastrophic failure. Pressouyre\cite{press1, press2} added to the theory that trapping plays a vital role in the kinetics of local hydrogen accumulation. McMahon and coworkers proposed the concept that co-segregation of hydrogen with impurity elements to the grain boundaries \cite{mcmahon} results in intergranular fracture.

Gerberich and co-workers attempted to combine the effects of localized plasticity and decohesion \cite{gerber1,gerber2}. In their numerical calculations, the stress enhancement due to dislocation activity is estimated to be of the order of 10-15 GPa which led to enhanced hydrogen accumulation, leading to local hydrogen concentration of about $50$ atomic percent. It should be noted that there is no direct experimental evidence for atomic hydrogen lowering interatomic forces of iron. Further, per literature models, the projected accumulation of hydrogen through his mechanism at the crack-tip is several orders of magnitude higher than what can be dissolved in bulk of the specimen.  }

\item{Hydrogen Enhanced Localized Plasticity (HELP): The hydrogen enhanced localized plasticity theory (HELP) proposes \cite{birnbaum1,birnbaum2,lynch} enhanced dislocation mobility in the presence of hydrogen. This theory considers that fracture occurs due to highly localized plasticity. The presence of hydrogen is proposed to decrease the resistance to slip due to reduced interaction stresses between dislocations containing hydrogen with the lattice and other obstacles, resulting in reduced flow stress. In situ TEM observations also showed higher mobility of dislocations in hydrogen atmosphere, resulting in shear localization in such regions\cite{birnbaum2} . However, the major limitation of the theory is that plasticity by itself cannot cause final failure without damage accumulation. Although proponents of this theory reported \cite{nagao, martin2011} formation of shallow nano-dimples in linepipe steels during hydrogen embrittlement, these dimples were still attributed to the HELP theory. Further, they did not provide a rational micromechanism by which the increased dislocation mobility and localization of plasticity led to the formation of nano-dimples.}

\item{Vacancy Model: According to the Vacancy theory proposed by Nagumo and co-workers\cite{nagumo1, nagumo2, nagumo3} , the main effect of hydrogen is to cause accumulation of excess vacancies when metals are deformed in the presence of hydrogen. Creation of a high density of vacancies, far exceeding the thermal equilibrium values is inferred to be a consequence of strong binding energy of vacancy with hydrogen. However, the evolution of vacancy accumulation to macroscopic fracture has not been established by this theory.}

\item{Nano-Void Coalescence (NVC): This theory proposed by Neeraj {\it et. al.} \cite{neeraj_actamat, li2015} attempts to connect the influence of hydrogen at atomic length scale to micro-scale damage accumulation leading to macroscopic failure. According to NVC model, hydrogen enhances localization of dislocation plasticity and stabilizes the excess vacancies generated by dislocation plasticity. The excess vacancies agglomerate aided by transport of vacancies by dislocation plasticity and diffusive processes to form nanovoids, which then coalesce into a macroscopic fracture pathway. While this is a recent contribution, it has gained significant interest in the hydrogen embrittlement research community because this work, for the first time, presented direct evidence for nano-voids on the fracture surface of hydrogen embrittled quasi-cleavage facets. Further, the NVC mechanism provides a rational micro-mechanical pathway to failure due to hydrogen. It is to be noted here, that the nanovoids in this context are assumed to be not filled with $H_2$ gas. Gas filled voids, also referred as bubbles, are at a different length scale (in the order of $100 \mu m$ - $500 \mu m$) compared to nanovoids and follow a different mechanism \cite{fischer}. The bubble blistering is not addressed here.}

\end{itemize}

\section{Numerical damage modeling literature}\label{model_lit}

There are numerous efforts in the literature to simulate the hydrogen embrittlement phenomenon at different length scales. Methods based on {\it ab initio} and atomistic simulations attempt to understand specific mechanisms of hydrogen interaction with metal lattice and defects. Continuum methods have been proposed to simulate overall macroscopic response due to the influence of hydrogen. Most of the continuum methods adopt a specific failure mechanism in their simulations. Some of the efforts at continuum scale also attempt to combine multiple theories into a single numerical model. Many of the efforts in the literature have reasonable success in predicting the time dependence of crack growth during hydrogen embrittlement. However, these models invoke unphysical conditions to induce failure such as reducing the strength of a macroscopic continuum fracture plane to zero as a function of hydrogen coverage on that plane. 

Gangloff  \cite{gangloff} developed analytical  models for estimating subcritical crack growth rate for high strength steels in the presence of hydrogen. Gangloff's work \cite{gangloff} and the references therein provide a good background for such efforts. These models were developed based on the decohesion theory and rely on local enhancement of hydrogen concentration of the order of $10^5$ to $10^6$ times the bulk concentration of few weight ppm to induce the failure. 

Serebrinsky {\it et. al.}  \cite{serebrinsky} reported a quantum-mechanically informed continuum model to simulate cohesive crack growth driven by HEDE theory. The model computes a coverage factor that quantifies the extent of hydrogen in the cohesive layer. The cohesive strength is degraded based on quantum-mechanics calculations as a function of the coverage factor.  Rimoli and Ortiz \cite{rimoli} reported a three-dimensional model for intergranular hydrogen embrittlement. The model degrades grain boundary cohesive strength based on coverage of hydrogen in a polycrystalline $4340$ steel double cantilever beam specimen. Their work predicts the crack initiation and growth to develop a crack growth rate curve and evaluate a limiting stress intensity factor. 

Novak {\it et. al} \cite{novak2} developed a micro-mechanically informed continuum model to predict fracture strength degradation in tempered martensitic high strength steel. In their work, hydrogen embrittlement is modeled as intergranular fracture along prior austenite grain boundaries. However, crack nucleation is due to hydrogen induced decohering of carbides based on HEDE theory. The hydrogen segregated to the grain boundaries reduces not only the work of adhesion, but also the attendant plastic work thus also accounting for the HELP mechanism. The model simulates degrading interface strength over a statistical distribution of carbides. The numerical calculations have been reported to be in good agreement with experimental results. These numerical models incorporate deformation coupled with hydrogen diffusion as originally developed by Sofronis {\it et. al.} \cite{novak2, rimoli,serebrinsky,sofronis1}. A more generalized, thermodynamically consistent thermo-mechanically coupled deformation and diffusion model is reported by Di Leo and Anand \cite{anand}. Anand {\it et. al.} also developed a coupled diffusion-deformation-damage model based on HELP mechanism \cite{anand2019}. They introduce a continuum damage variable to track the state of each material point from intact to fully fractured condition. In their model, the limiting case of zero hydrogen lattice concentration leads to localized necking and does not account for ductile fracture through the classical microvoid  growth and coalescence mechanism. Wasim {\it et. al.} \cite{wasim1, wasim2} developed a fracture toughness degradation model based on their experiments and in accordance with HELP and HEDE theories. The synergistic effects between HELP and HEDE were also adopted for fatigue life of BCC steel for hydrogen embrittlement using an ab-initio based Unified Mechanics Theory \cite{weilee}. Some of the recent efforts also adopted phase field methods for modeling hydrogen embrittlement. The phase field method approximates the sharp crack morphology in material with a diffusive crack zone that transitions between intact and fully separated material. Hydrogen transport is coupled with displacement and phase field, adopting HELP+HEDE mechanism by Huang \cite{huang}. Recent literature also includes cohesive zone modelling for hydrogen assisted fatigue crack growth using strain gradient plasticity \cite{sousa}, phase-field method for dissolution-driven stress corrosion cracking and hydrogen embrittlement based on J2 plasticity \cite{cui} and combined phase field and cohesive zone formulation for hydrogen embrittlement in polycrystalline microstructure that explicitly captures fracture interaction with granular structure in the presence of hydrogen under HEDE mechanism \cite{gonzalez}.

Sofronis {\it et. al.} \cite{sofronis_aravas} adopted the HELP mechanism to simulate hydrogen induced plastic instability. While HELP itself is not a fracture mechanism, the effect of hydrogen is modeled through hydrogen induced volume dilatation and reduction in local flow stress, leading to plastic flow localization. The flow stress expression is modified to simulate reducing yield stress with increasing hydrogen concentration.  

Nagumo adopted his vacancy theory into a numerical model that simulates effect of hydrogen as higher initial void volume fraction on a ductile material model \cite{nagumo1, nagumo3} . By simulating fracture resistance curves, Nagumo attempted to support the vacancy model that hydrogen increases the strain-induced vacancies thus enhancing void formation, followed by plastic instability and decrease of crack growth resistance.

In the area of atomistic modeling efforts, Song and Curtin performed atomistic simulations of hydrogen in BCC Iron\cite{curtin1, curtin2} , to conclude that hydrogen forms Cottrell atmosphere around moving dislocations, resisting dislocation motion through solute drag. However, they find no evidence of HELP mechanism in their studies. In other studies they have performed atomistic simulation to propose that hydrogen accumulation at crack tips can reach very high concentrations to form hydrides. The hydride is proposed to inhibit dislocation emission from crack tips and instead promote brittle cleavage fracture. The weakness of the model is that there is no experimental evidence for formation of hydrides in steels or Ni based alloys. Further, the role of dislocation emission from crack-tips as the controlling mechanism for ductility in metals at quasi-static loading rates is also not clear since metals typically have a very high density of pre-existing mobile dislocations. Further, there is experimental evidence that significant localized plasticity is associated with hydrogen embrittlement. 

Ju Li and his group have performed a series of modeling studies\cite{juli} at different length and time scales to understand the mechanisms of hydrogen enhanced excess vacancy accumulation and their agglomeration to form nano-voids. Using large-scale Molecular Dynamics (MD) simulations it has been shown that plastic deformation in the presence of hydrogen did lead to excess vacancy accumulation reaching concentrations ($\sim C_v = 10^{-3}$) that are comparable to vacancy concentrations found near melting point. Further, MD simulations also showed that the hydrogen-vacancy complex is quite stable in that it cannot be swept away by typical dislocation processes and the migration energy is higher (0.76 eV) and so cannot be easily annihilated by diffusive processes. Further, they showed that typical dislocation processes can assist in transporting vacancies to hydrogen-vacancy complex and grow them by a displacive process. Using Kinetic Monte-Carlo (kMC) and Cluster Dynamics modeling it is shown that hydrogen-vacancy complex growth into proto nano-voids can occur by conventional diffusive processes as well. In summary, their work provides additional atomistic and mechanistic basis for the NVC theory. 

The present work aims to develop a continuum damage model to simulate hydrogen embrittlement based on the NVC mechanism. The rest of the article presents the new damage model and experimental studies for validation. Section \ref{damage_model} discusses the proposed damage model and its coupling with hydrogen transport process. Section \ref{development} discusses the model development and it also describes sensitivity studies on the model parameters in predicting loss of ductility by the damage model. Section \ref{example} discusses tensile tests on round bar specimen electrochemically precharged with hydrogen and the corresponding numerical simulation of these tests. The article concludes with discussion of results and a summary in sections \ref{discussion} and \ref{summary} respectively.

\section{Continuum Damage Model}\label{damage_model}

As evident from the hydrogen embrittlement modeling literature, the mathematical foundation of hydrogen damage theory relies on a specific fracture mechanism. In this work, the continuum damage model developed relies on the Nano-Void Coalescence (NVC) theory. Considering that under normal circumstances (in the absence of hydrogen) steels exhibit ductile behavior, the basic continuum methodology for simulating such response needs to consider void evolution in a ductile solid with non-deforming inclusions. Typically ductile failure in metals occurs by a Micro-Void Coalescence (MVC) mechanism that initiates failure by void nucleation through interfacial debonding or cracking of inclusions, void growth due to local triaxial stress state and finally void coalescence leading to macroscopic fracture. Since the NVC theory has close similarities to micro-void coalescence theory, the mathematical foundation for the proposed damage model is developed as a modified Gurson model for ductile material. 

\subsection{Gurson damage model}
Early micromechanical studies on ductile failure of metals\cite{rice} showed that rate of growth of voids is strongly dependent on hydrostatic tension and void coalescence is promoted by higher triaxial tension. For a ductile material containing a volume fraction {\it f} of voids, Gurson has suggested a plastic yield condition of the form $ \Phi( q, p, \sigma_0, f)=0$ where, $q$ and $p$ are Mises stress and hydrostatic stress respectively, $\sigma_0$ is equivalent tensile flow stress. Based on a spherical void in a spherical matrix medium, the yield condition is derived as:

\begin{equation}
\Phi \equiv \left( \frac{q}{\sigma_0} \right) ^2 + 2q_1f ~ \cosh \left( \frac{-3q_2 p}{2\sigma_0} \right) -(1+q_3 ~ f^2) = 0
\label{gurson_yieldfn}
\end{equation}
 where-in voids are assumed to be homogeneously distributed. Micro-mechanical studies of Gurson \cite{gurson} and Tvergaard \cite{tvergaard, tvergaard_mvc} led to deducing the constants, $q_1 = 1.5$, $q_2 = 1.0$ and $q_3 = q_1^2$ as widely applicable for metals.

The above yield condition reduces to von Mises yield function($q-\sigma_0=0$) for void fraction $f=0$. The material microstructure is assumed to consist of plastic incompressible matrix and brittle inclusions. This model includes the effect of hydrostatic stress on plastic flow. Hence it allows for volume change in the matrix of the medium unlike von Mises theory that is volume preserving. The Gurson model is an associative plasticity law and consequently the macroscopic plastic strain increment is given by the following normality rule.

\begin{equation}
d \epsilon ^{p} = d \Lambda \frac{\partial \Phi}{\partial \sigma}
\end{equation}
where $\Lambda$ is a plastic parameter.

The void volume fraction is a state variable for this constitutive model and its variation is governed by an evolution law. The change in void volume fraction during a load increment is considered to be a combination of 

\begin{itemize}
\item nucleation of new voids by cracking or decohesion of inclusions, $df_{nucleation}$
\item growth of existing  voids due to hydrostatic stress, $df_{growth}$
\end{itemize}

Hence, the void growth law is given by 

\begin{equation}
df = df_{nucleation} + df_{growth}
\end{equation}

The nucleation is considered to be controlled by plastic strain given by 
\begin{equation}
df_{nucleation} = A ~ d \bar{\epsilon^{p}}
\label{df-carbides}
\end{equation}
where the parameter $A$ is driven by normal distribution of nucleation strain\cite{chu} with mean value $\epsilon_N$ and standard deviation $S_N$.
\begin{equation}
A = \frac{f_N}{S_N \sqrt{2\pi}} exp\left[-\frac{1}{2} \left(  \frac{\bar{\epsilon^{p}} - \epsilon_N}{S_N}  \right)^2    \right]
\label{df-carbides2}
\end{equation}
where $f_N$ is volume fraction of nucleating particles in the material. 

The change in void volume resulting from growth of voids is given by
\begin{equation}
df_{growth} = (1-f) d\epsilon^p : {\bf I} \quad \equiv  \quad (1-f)d\epsilon^p_{kk}
\end{equation}

From the yield condition, one can deduce that the yield surface shrinks to a point at $f=\frac{1}{q_1}$. Since this is an unrealistic value, a number of studies found that coalescence occurs at a stage when the size of void grows to be close to the spacing between them. At this stage, local failure occurs by development of slip bands between the voids  in the material. This microscopic necking phenomena has been incorporated as the effect of coalescence by Tvergaard and Needleman \cite{tverg_needleman} through a void acceleration function.

\subsection{Modified Gurson model for NVC based hydrogen embrittlement}

The NVC mechanism is translated into a damage model using the above described Gurson model with two additional contributions from NVC theory. (i) hydrogen enhanced localized plasticity and (ii) hydrogen enhanced void nucleation. The contributions of these two effects to the proposed damage model are described below.

\subsubsection{Hydrogen enhanced localized plasticity}
As observed in NVC mechanism, hydrogen enhances localization of dislocation plasticity. This translates to abnormal (high) plastic strain for a given stress state in those regions of the sample where there is an enhanced concentration of hydrogen. This is often represented by an expression for flow stress dependent on hydrogen concentration, in addition to plastic strain. There are multiple forms of such expression in literature to represent HELP process. For instance, Sofronis {\it et. al.}\cite{liang, novak} proposed a form with trapped hydrogen concentration as 
\begin{equation}
\sigma_{flow} (\bar{\epsilon^p}, C_H) =  \sigma_o  f(\bar{\epsilon^p}) \left[ (\xi -1)\frac{C_H}{C_o}+1   \right]    \nonumber
\end{equation}
where $\sigma_o  f(\bar{\epsilon^p})$ is the flow stress in the absence of hydrogen as a function of effective plastic strain, $\bar{\epsilon^p}$, $\xi \le 1$ is a parameter denoting softening, $\xi >1$ denotes hardening, and $C_H/C_o$ is the normalized hydrogen concentration in trapping sites. The expression $ f(\bar{\epsilon^p})$ can take a typical power law expression such as $(1+ \frac{\bar{\epsilon^p}}{\epsilon_o})^n$. In the limiting condition of $\xi =0$ with $C_H = C_o$, the flow stress reaches a zero value. A modified version of this expression has also been reported by Sofronis and coworkers \cite{petros_ductile} that includes another parameter to limit the reduction of flow stress due to hydrogen. In the modified expression, the limiting value of extreme softening is defined as $\eta \sigma_0$ where $\eta$ is a model parameter.

Similarly Murakami, and coworkers \cite{lijun} proposed a softening model using the expression
\begin{equation}
\sigma_{flow} (\bar{\epsilon^p}, C_H) = \hat{ \sigma_o} (\bar{\epsilon^p}) \left[ (1+ \xi C_H ) \left(1+  \frac{C_H}{C_c} \right)^n   \right] \nonumber
\end{equation}

The damage model in this work consists of a similar phenomenological expression for flow stress as a function of hydrogen exposure given by,
\begin{equation}
\sigma_{flow} (\bar{\epsilon^p}, C_H) = \sigma_o \left[ 1+ \frac{\bar{\epsilon^p}}{\epsilon_o} \left(1 - \frac{C_H}{C_c} \right)   \right]^n 
\label{h-softening}
\end{equation}
where, $C_H$ is the amount of reversible hydrogen concentration,  $C_c$ is critical amount of hydrogen concentration that results in maximum softening and $n$ is the hardening exponent. In the above functional form for flow stress, as $C_H  \rightarrow  0$, the flow stress reaches the regular power law for hardening. At the other extreme, as  $C_H  \rightarrow  C_c$, the flow stress reaches $ \sigma_o $ and does not depend on $\epsilon^p$ anymore representing perfectly plastic deformation. Since the expression captures solely hydrogen enhanced softening and not damage, the yield surface is not expected to shrink below the radius $\sigma_o$, at $C_H = C_c$. Note that the proposed expression contains a hydrogen related material parameter, $C_c$. 

\subsubsection{Hydrogen enhanced void nucleation}
As described in the work of Neeraj and coworkers\cite{neeraj_actamat}, the excess vacancies generated due to dislocation plasticity are stabilized by the presence of hydrogen and tend to agglomerate into proto nano-voids that eventually grow and coalesce to form either nano or micro-voids. This phenomenon is captured as an additional void nucleation process in the damage model. It should be emphasized here, that the excess vacancies are generated and accumulated due to dislocation plasticity in the presence of hydrogen. Hence the new void evolution law in the presence of hydrogen is given by 

\begin{equation}
df = df_{nucleation} + df_{nucleation,H} + df_{growth}
\label{df-law}
\end{equation}

where $df_{nucleation,H}$ is the additional nucleation due to the presence of hydrogen and is given by

\begin{equation}
df_{nucleation,H} = B ~ d \bar{\epsilon^{p}}  ~ C_H
\label{h-nucleation}
\end{equation}

where $B$ is a parameter (units $m^3/mol$) for hydrogen induced nucleation of voids in the material. The expression includes the incremental effective plastic strain to represent the key role of dislocation plasticity in the agglomeration of vacancies into voids. The parameter $B$ quantifies the interactions between hydrogen vacancy complexes and dislocation plasticity. Hence it is denoted as the $H-V$ complex coefficient in this damage model.

From the above proposed modifications to the Gurson model, the presence of hydrogen enhances localized plasticity through the reduction of the flow stress reaching to a minimum of rigid, perfectly plastic condition at the extremity. Hydrogen also enhances generation of new voids $df_{nucleation,H}$ that is dependent on the incremental effective plastic strain and (local) hydrogen concentration. It is also important to note that the localized plasticity does not directly affect the damage parameters of the material while the excess vacancy stabilization directly enhances damage through accelerated void nucleation. Also, while the proposed model incorporates void nucleation due to hydrogen, void growth in the presence of hydrogen is assumed to follow the classical Gurson theory.

\subsection{Coupled damage and hydrogen transport}

The mechanical deformation of a structure exposed to hydrogen is simulated with two coupled processes as proposed by Sofronis and coworkers \cite{sofronis1, petros_ductile,dadfarnia} . The deformation problem involves solving the quasi-static equilibrium equation under large deformation and the hydrogen transport problem involves hydrogen mass balance in the lattice and traps. 

The deformation of the structure is affected by lattice dilation rate due to dissolved hydrogen as described by, 
\begin{equation}
\epsilon^H_{ij} = \left[ \frac{ \frac{\Delta v}{3 \Omega}  }{\frac{N_L}{N_A} + C_H \frac{\Delta v}{3 \Omega} }  \right] \Delta C_H \ \delta_{ij}
\label{hstrain-eqn}
\end{equation}
where, $\Delta C_H$ is incremental hydrogen concentration between two successive time steps (in $moles/m^3$), $\Delta v$ is volume increase of host lattice per mole of hydrogen introduced, $\Omega$ is the molar volume of host metal, $N_L$ is the number of metal lattice sites, $N_A$ is Avogadro number. With $V_H$ as partial molar volume of H and $V_M$ as molar volume of host lattice, $\Delta v = V_H/N_A$ and $\Omega=V_M/N_A$.

Consequently, the strain decomposition is written as 
\begin{equation}
\epsilon = \epsilon^e + \epsilon^p + \epsilon^H
\end{equation}

The evolution of $\epsilon^p$ is guided by the Gurson plasticity model and the evolution of lattice dilation strain $\epsilon^H$ calculated from equation \ref{hstrain-eqn} is guided by hydrogen transport model.

The hydrogen transport mechanism is controlled by stress driven hydrogen diffusion phenomenon as proposed by Sofronis and coworkers. The mathematical formalism of hydrogen transport in metals developed by Sofronis and coworkers has also been implemented by several other authors in the literature\cite{serebrinsky} . In this work the trap mediated hydrogen diffusion theory is implemented. The equilibrium between the occupancy of lattice and trap sites \cite{oriani} , is given by Oriani's equation,
\begin{equation}
\frac{\theta_T}{1-\theta_T}=\frac{\theta_L}{1-\theta_L} exp \left( -\frac{W_B}{RT}  \right)
\label{oriani-eqn}
\end{equation}

where, $\theta_L$, $\theta_T$ are occupancies in lattice and trap respectively and $W_B$ (as a negative quantity) is the binding energy of the trap.

Relating the occupancy with hydrogen concentration with the expressions $C_T=\theta_T \alpha N_T$ and $C_L=\theta_L \beta N_L$ in the above equation gives the relation between lattice and trap concentrations as
\begin{equation}
C_T = \frac{\alpha N_T K_T C_L}{\beta N_L + (K_T-1)C_L} \qquad \mbox{where} \quad K_T =  exp \left( -\frac{W_B}{RT}  \right)
\end{equation}

where $\beta$ is the number of interstitial lattice sites per solvent atom and $N_L$ is the number of solvent atoms per unit volume,  $\alpha$ is the number of sites per trap and $N_T$ is the number of traps per unit volume. The above expression connects the effect of traps on diffusion of lattice hydrogen, based on effective diffusivity equation given by

\begin{equation}
D_{eff} = \frac{D}{1+ \frac{\partial C_T}{\partial C_L} }
\end{equation}

where $D$ is diffusivity of hydrogen in pure iron and $D_{eff}$ is the effective diffusivity due to the presence of traps.

The expression $\frac{\partial C_T}{\partial C_L}$ is given by
\begin{equation}
\frac{\partial C_T}{\partial C_L} = \frac{\alpha N_T K_T \beta N_L}{ [\beta N_L + (K_T-1)C_L]^2}
\label{ctcleqn}
\end{equation}

The diffusion equation for hydrogen transport in lattice and traps is given by 

\begin{equation}
\frac{D}{D_{eff}}\frac{\partial C_L}{\partial t} + \alpha \theta_T \frac{\partial N_T}{\partial \epsilon^p}  \frac{\partial \epsilon^p}{\partial t} + J_{i,i} = 0
\label{h-transport}
\end{equation}

where, the second term is associated with dislocation traps and it assumes that $N_T$ is a function of equivalent plastic strain ($\epsilon^p$) for dislocation traps ($N_T = N_T(\epsilon^p)$). The deformation influenced flux is given by 

\begin{equation}
J_i = -DC_{L,i} + \frac{DV_H C_L}{3RT}\sigma_{kk,i}
\end{equation}

The above equation \ref{h-transport} is solved in a finite element framework on the same mesh used for the deformation and damage simulation.

In the coupled problem, the deformation and transport problems are solved in a staggered scheme. The quasi-static equilibrium is attained at a specified hydrogen distribution and the transport problem is solved based on the equilibrium hydrostatic stress. The deformation equilibrium equations are solved using ABAQUS finite element software \cite{abaqus}, through a user material routine (UMAT). The Gurson constitutive relation is a nonlinear stress-strain response and needs special attention for stress update. The numerical algorithm proposed by Aravas \cite{aravas} for stress update in pressure dependent plasticity model is implemented in this work. The equilibrium equations are solved in implicit scheme and hence need a consistent tangent modulus for the Newton-Jacobian. The consistent tangent modulus proposed by Zhang \cite{zhang} has been demonstrated to be a robust method and is implemented in this work. The transport problem is also solved in an implicit time stepping scheme and is a linear problem. The solution of transport problem is primarily obtained in $C_L$, lattice concentrations, that is used to update the trap concentrations ($C_T$). The sensitivity of the proposed damage model to its hydrogen based damage parameters, namely (i) critical hydrogen concentration $C_c$ based softening and (ii) vacancy based void nucleation parameter $B$ are discussed in the next section.

\section{Damage model development}\label{development}
The material addressed in this work is a X65 grade linepipe steel. The basic mechanical properties vary between different steel makers and even between different production batches. The nominal mechanical properties can be stated as $\sigma_o=420MPa$, Ultimate Tensile Strength (UTS)$ = 510MPa$ and tensile elongation $e_f =18\%$. In the current work, the tensile test is taken as the methodology to characterize the mechanical behavior. The tensile sample is characterized by a gage length of $2.54cm (1 inch)$ and cross-section diamater $0.3175cm (1/8 inch)$, with applied tensile strain rate of $10^{-5} /sec$ until failure. While steels contain several types of traps such as grain boundaries and carbide interfaces, the material here is assumed to contain predominantly dislocation traps and these traps evolve with local effective plastic strain $\bar{\epsilon^p}$. The dislocation traps are assumed to have a binding energy $20.2 kJ/mol$ \cite{dadfarnia} . The dislocation trap density $N_T^{(D)}$ increases with plastic strain $\bar{\epsilon^p}$. Assuming one trap per atomic plane threaded by a dislocation, 
\begin{equation}
N_T^{(D)} = \sqrt{2}\rho/d
\end{equation}
where $\rho$ is the dislocation density and $d$ is lattice parameter. The dislocation density in this work is given by \cite{dadfarnia}
\begin{eqnarray}
\rho =\left\{ \begin{array}{ccc} \rho_{o} + 2\gamma\bar{\epsilon^p}  &  \mbox{for} & \bar{\epsilon^p} \leq 0.5 \\  \mbox{const.}  &   \mbox{for} & \bar{\epsilon^p} > 0.5 \end{array} \right.
\end{eqnarray}

where $\rho_o = 10^{10} length/m^2$ and $\gamma = 10^{16} length/m^2$. 

The basic mechanical properties and assumed trap properties are listed in table \ref{properties}.

\begin{table}[htbp] 
\begin{center}
\begin{tabular}{ c c c } \hline
Property & Symbol & Value \\ \hline
Young's modulus                & $E$          & $200 GPa$ \\
Poisson's ratio                & $\nu$        & $0.3$        \\
Yield stress                   & $\sigma_0$   & $420 MPa$    \\
Number of interstitial lattice sites per host atom   &  $\beta$      & $1$          \\
Molar volume of the host lattice& $V_M$       & $7.116 cm^3/mol$ \\
Partial molar volume of H      &  $V_H$       & $2cm^3/mol$  \\
Diffusion coefficient          &  $D$         & $1.271 \times 10^{-8} m^2/s$ \\
Dislocation binding energy     & $W_B^{(D)}  $& $20.2 kJ/mol$ \\
Number of trapping sites per dislocation & $\alpha^{(D)}$ & $1$ \\
Temperature              & $T $         & $300 K$ \\
Lattice parameter              & $d $         & $2.86 A^o$ \\  \hline
\end{tabular}
\end{center}
\caption{Material properties for the Steel studied in this work}
\label{properties} 
\end{table}

\subsection{Calibration}
The basic (hydrogen free) Gurson parameters for this material are taken as $q_1=1.5$, $q_2 = 1.0$, $q_3 = q_1^2$. Considering minimal initial voids in the as received material, the initial void volume fraction is assumed to be $0.0001$. The void nucleation parameters are taken as $f_N = 0.05$, $\epsilon_N = 1.0$ and $S_N = 0.1$.  A sensitivity analysis to study the variation of tensile stress-strain response with the nucleation parameters led to the inference that these are a suitable set of parameters to model the reference material properties. An abundance of literature\cite{tvergaard_mvc} is available in regards to the parameters for the Gurson model that support the range of these parameters. The detailed sensitivity analysis for basic (hydrogen free Gurson model) parameters is not presented here.

\begin{figure}
\begin{center}
\includegraphics[scale=0.25]{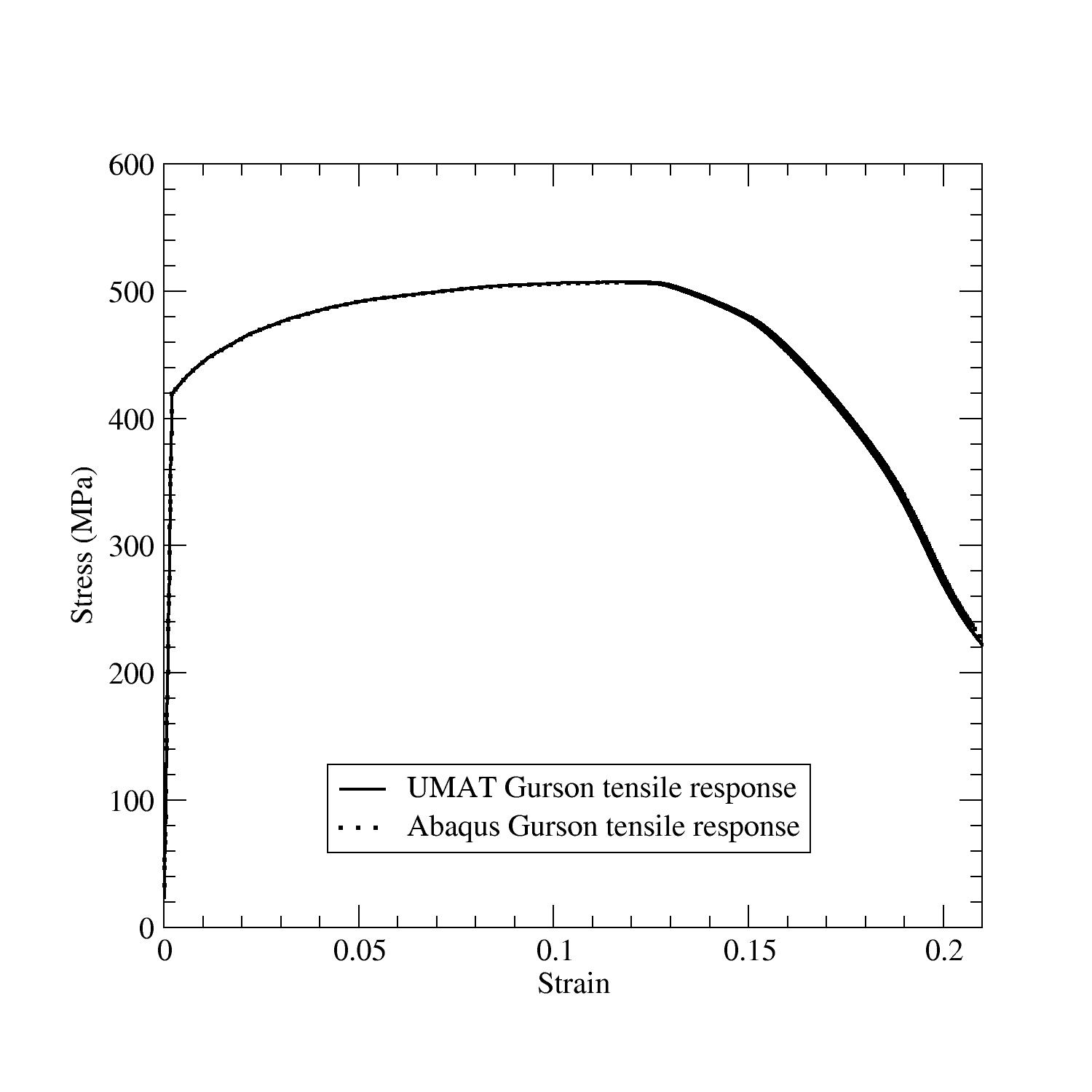}
\end{center}
\caption{Numerical tensile stress-strain reponse for round bar specimen using Gurson user material (UMAT) developed in this work and Abaqus in-built Gurson constitutive model.}
\label{fig-validation}
\end{figure}

The hardening curve without hydrogen is taken as $\sigma_f = \sigma_o (1+\bar{\epsilon^p}/\epsilon_o)^n$ with $\sigma_o = 420$, $\epsilon_o = 0.012$ and  $n=0.13$. It is to be noted here that the tensile stress strain curve is sensitive to the hardening parameters and the parameters assumed here may not be a unique combination to produce the desired stress-strain curve. However, the choice of parameters is typical of those used for steels. A round bar geometry with three orthogonal planes of symmetry is generated for the tensile specimen. A defect is simulated at the center of symmetry line of the cross-section (middle of the gage), by reducing the radius of specimen \cite{aravas} in this plane by $\Delta R = 0.005R$ or $0.5\%$. This imperfection naturally occurs in a material due to the deformation microstructure and the crystallographic nature of slip.  It is introduced in the simulation here to provide a necking location in an otherwise homogeneous continuum. This ensures that the necking of the specimen is triggered not by numerical instabilities, but by the intentionally built-in defect. The obtained stress-strain curve with the above material parameters is shown in figure \ref{fig-validation}, denoted as {\it UMAT Gurson tensile reponse}. As evident from the tensile response, it closely represents the above mentioned basic (hydrogen free) mechanical properties of the material i.e., initial yield stress, UTS and elongation to failure.

\subsection{Validation studies}
The basic Gurson model is available as a material model in ABAQUS software. However, the modified Gurson model for hydrogen embrittlement (hereafter called NVC damage model) needs a unique material model development. Hence the proposed damage model in this work has been developed in the form of ABAQUS user material routine. Hence the first step in the process of testing such a development is to ensure the integrity of NVC damage model calculations without the influence of hydrogen. The tensile response generated from the above calibration is also calculated using the ABAQUS in-built Gurson model using the same material parameters. As shown in figure \ref{fig-validation}, the tensile stress-strain response of NVC damage model developed in this work matches very well with the ABAQUS constitutive model. The coupled deformation and diffusion problem without damage is bench-marked with the results of Dadfarnia\cite{dadfarnia} and coworkers. The detailed validation studies with the published work are not discussed in this article.

\subsection{Sensitivity analysis}

\begin{figure}
\begin{center}
\includegraphics[scale=0.25]{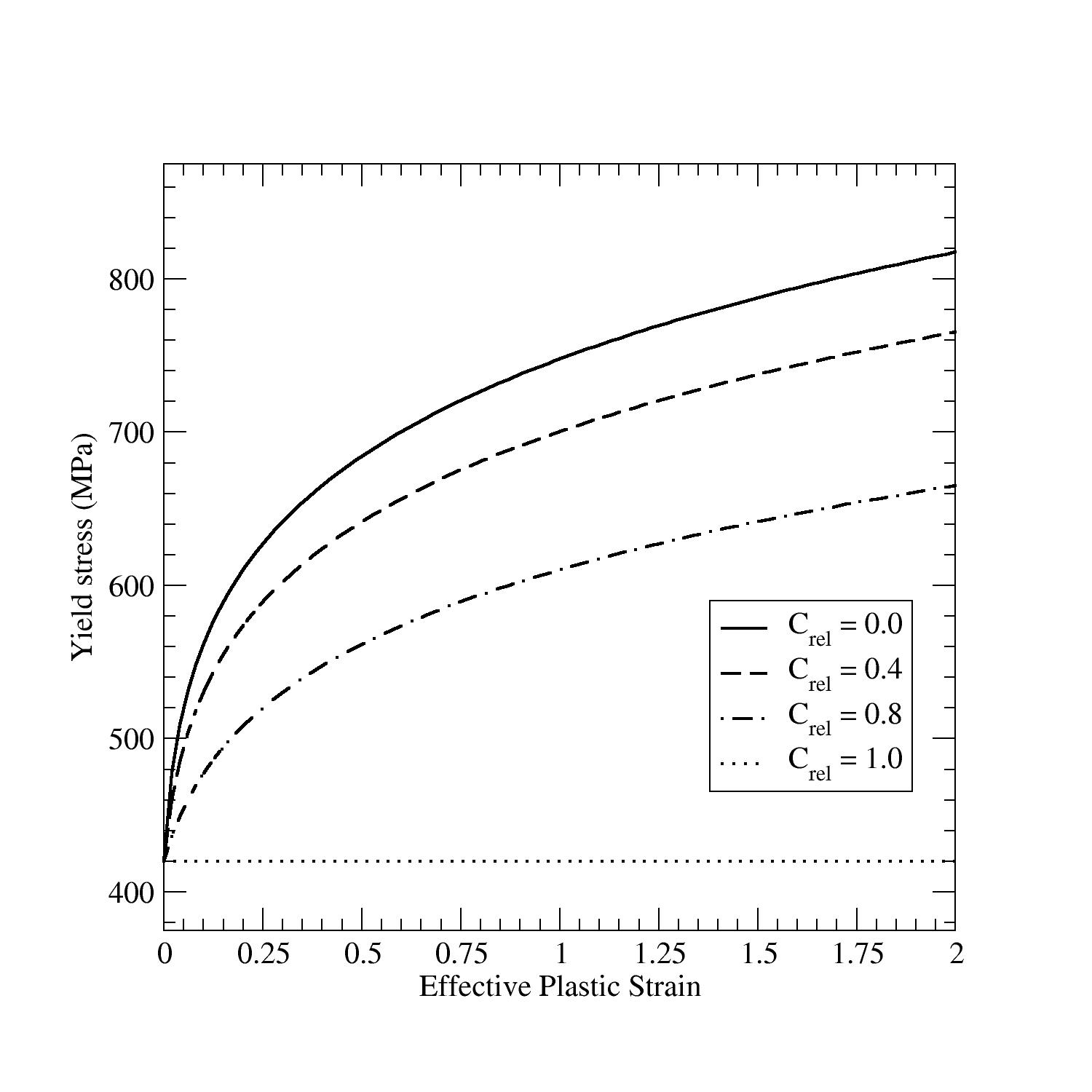}
\end{center}
\caption{Strain-hardening for the model material with various exposures of hydrogen. $C_{rel}=0.0$ indicates no hydrogen and $C_{rel}$ indicates critical amount of hydrogen. With the modified hardening law, the material softens as the exposure of hydrogen increases.}
\label{fig-soft_law}
\end{figure}

The contribution of hydrogen to damage accumulation and the sensitivity of model parameters to the macroscopic response are presented next. First, the effect of hydrogen enhanced softening is analyzed. From the hydrogen based hardening law in equation \ref{h-softening}, for a constant $C_c$, as hydrogen exposure $C_H$ increases the flow stress is expected to decrease below its normal value (without hydrogen). The new material parameter that is introduced to quantify the influence of hydrogen is $C_c$, the critical amount of hydrogen that can result in high dislocation plasticity. The ratio, $\frac{C_H}{C_c}$ is hereby referred as $C_{rel}$, bounded by $0$ and $1.0$. The hardening law at any point in the material responds to the local hydrogen concentration based on equation \ref{h-softening}. The sensitivity of hardening law to relative hydrogen concentration $C_{rel}$ is plotted in figure \ref{fig-soft_law}. As evident from the plot, the material responds with normal hardening in the absence of hydrogen and reaches the extreme of no hardening at $C_{rel} = 1.0$. Since the effect of hydrogen is in the form of a power-law expression, the flow stress varies nonlinearly as $C_{rel}$ reaches $1.0$.

\begin{figure}
\begin{center}
\includegraphics[scale=0.25]{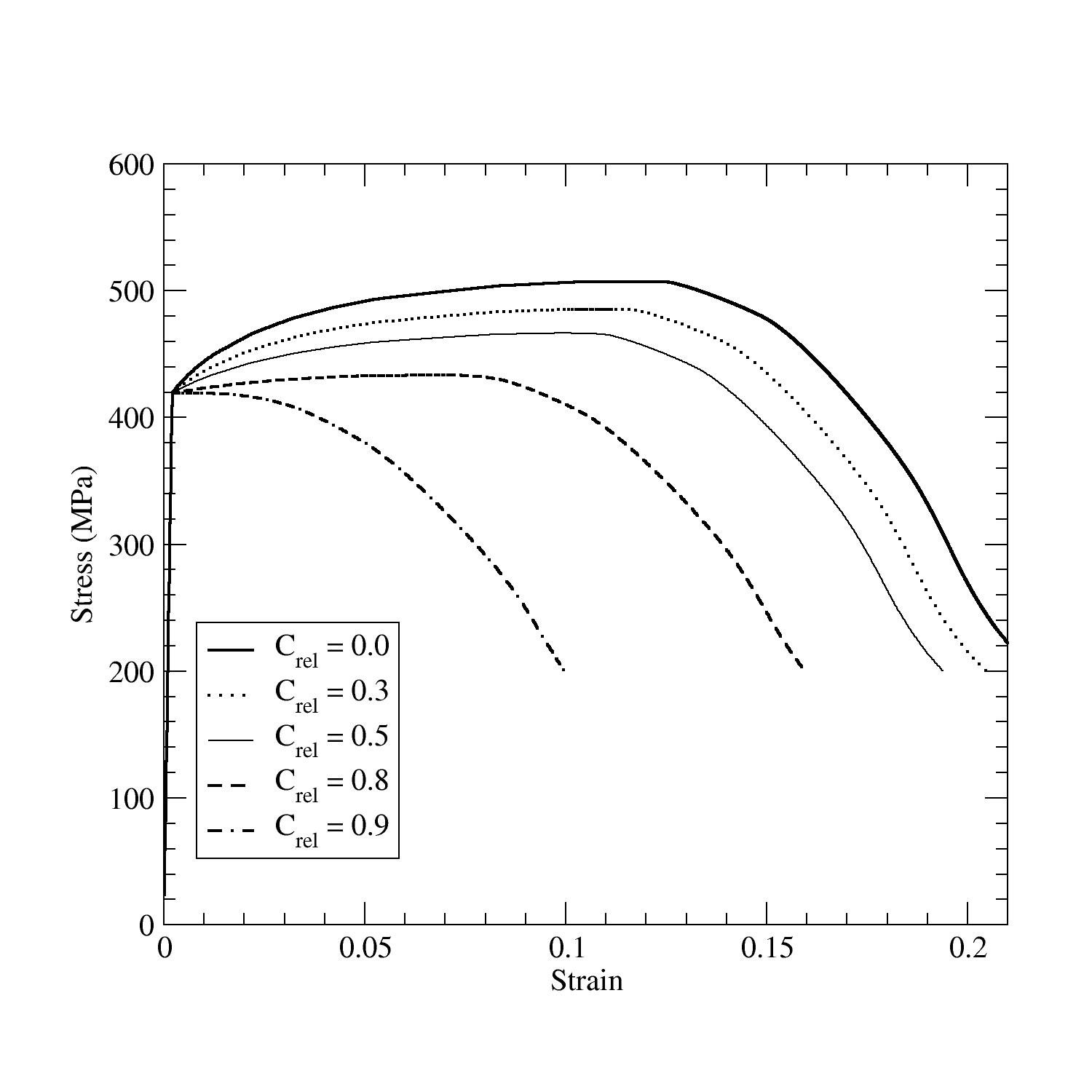}
\end{center}
\caption{Tensile stress-strain response of round bar specimen with the effect of hydrogen induced softening.}
\label{fig-soft}
\end{figure}

In the following sensitivity analysis, the tensile round bar specimen is precharged with a total of $4 ppm(wt.) $ or $3.15 \times 10^{-8} mol/mm^3$ of hydrogen that is homogeneously distributed and the value of $C_c$ is varied. In effect, the normalized quantity $C_{rel}$ attains various values between $0$ and $1.0$. A number of tensile tests are simulated with varying values of $C_{rel}$ and the macroscopic stress-strain response is plotted in figure \ref{fig-soft}. It is to be noted at this point, that the hydrogen diffusion calculations do not result in much change of hydrogen distribution along the necking region. However, there is significant plastic strain localization. The combination of localized strain and hydrogen presence plays a predominant role in the overall softening behavior. During these simulations, the effect of hydrogen induced nucleation of voids is disabled to focus on the contribution of HELP mechanism. 

The response at $C_{rel}=0.0$ refers to in-air tensile properties. The increase in hydrogen is represented as the increase in $C_{rel}$, and accordingly modifies the corresponding local hardening curve(at integration point) in the material. As evident from this plot, as $C_{rel}$ increases, the material tends to have higher impact of hydrogen, resulting in lower flow stress and increasing loss of strength. It is to be noted here that the change in tensile response comes due to altered hardening curve, shrinking the yield surface for a material point. The overall response can be characterized to be softening macroscopic behavior as the exposure of hydrogen increases. Also the tensile response is more sensitive to $C_{rel}$ as it reaches closer to $1.0$.

The response also shows significant loss in tensile strength(UTS) of the material. However, the loss of UTS with increasing hydrogen is not observed in experimental tensile tests. The trend observed in numerical simulations may be explained through two reasons: (i) The values of $C_{rel}$ in reality may be in the order of $0.01$ or smaller resulting in insignificant loss of UTS as predicted by the model; (ii) The enhanced plasticity is localized to a much smaller region/volume that is not captured by continuum analyses.(The continuum analyses may be over-estimating the region of softening due to mesh dimensions)

The contribution of hydrogen stabilized vacancy agglomeration is represented in the model through equation \ref{h-nucleation}. This expression involves the influence of plastic strain and hydrogen concentration, together with a material dependent parameter, $B$. The $H-V$ complex coefficient $B$ represents the influence of hydrogen to enhance void nucleation. The contribution of hydrogen induced nucleation is expected to increase significantly as the value of $B$ increases. It is a continuum parameter representing a number of atomic scale interactions, resulting in enhanced void nucleation. 

The damage in the material is quantified by void fraction that evolves based on equation \ref{df-law}. In the damage evolution law, the void nucleation comprises of two components. The primary component is given by equation \ref{df-carbides} that defines the nucleation of voids due to splitting of carbides or their interfacial debonding. The second component is the contribution of hydrogen through $H-V$ complex stabilization, given by equation \ref{h-nucleation}.

\begin{figure}
\begin{center}
\includegraphics[scale=0.25]{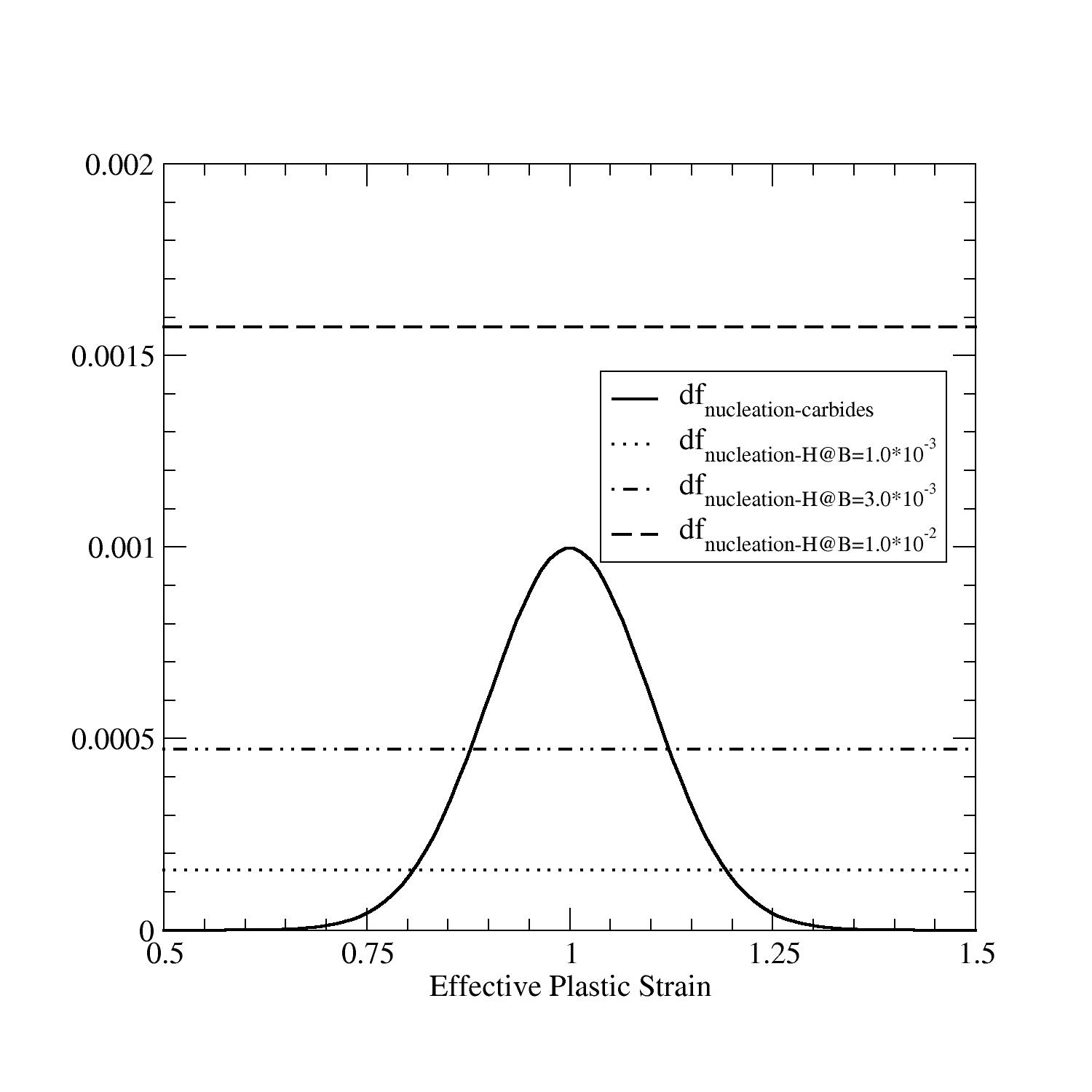}
\end{center}
\caption{Variation of void nucleation df from breaking carbides and hydrogen with effective plastic strain. Hydrogen contribution is plotted for various values of $H-V$ complex parameter $B$. Y-axis represents the increment of void nucleation. }
\label{fig-nuc_law}
\end{figure}

The relative contribution of these two nucleation sources is calculated by varying effective plastic strain $\bar{\epsilon^p}$ at constant incremental plastic strain $d \bar{\epsilon^p} = 0.005$. The values are plotted in figure \ref{fig-nuc_law}, that shows different dominant regimes of nucleation sources for a range of effective plastic strain ($\bar{\epsilon^p}$). The contribution from carbides is from equations \label{df-carbides} and \label{df-carbides2}, and the contribution from hydrogen is from equation \label{h-nucleation}.  The figure shows carbides based nucleation to be symmetric and maximum about $\epsilon_N = 1.0$. This is due to the assumed statistical distribution of carbides that split near $\epsilon_N = 1.0$. The plot also shows constant contribution of hydrogen to nucleation. This is because $df_{nucleation-H}$ varies with $d \bar{\epsilon^p}$ but does not depend on $\bar{\epsilon^p}$.

\begin{figure}
\begin{center}
\includegraphics[scale=0.25]{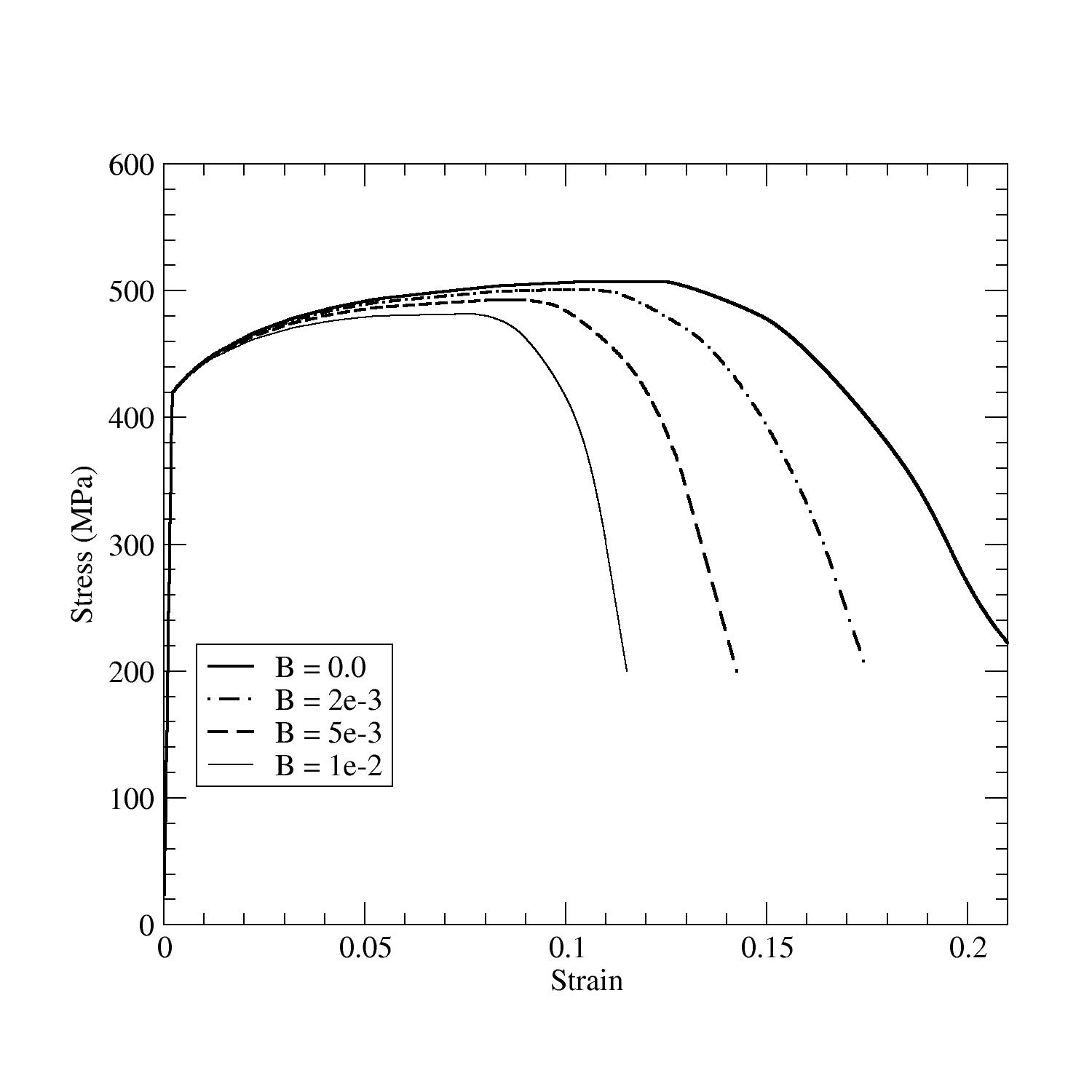}
\end{center}
\caption{Tensile stress-strain response of round bar specimen with the effect of hydrogen induced void nucleation.}
\label{fig-nuc}
\end{figure}

The sensitivity of macroscopic tensile response for the round bar tensile specimen is analyzed with varying $H-V$ complex coefficient. The response for a range of $B$ values is plotted in figure \ref{fig-nuc}. In these simulations, the hydrogen induced softening is disabled to emphasize the nucleation component. As evident from the trend in tensile stress-strain response, the enhanced nucleation reduces the tensile elongation significantly. Also important to note is that the loss in tensile strength (UTS) is not significant ($<20MPa$) for the range that is tested, which is consistent with experiments. 

In summary, the sensitivity analyses show that the tensile response is sensitive to both hydrogen induced softening and nucleation. Whereas the softening contribution reduced the elongation to failure and UTS significantly, the nucleation terms impacted primarily the elongation to failure. As seen in the next section, a combination of these effects can produce a realistic prediction of tensile stress-strain response with the influence of hydrogen. A set of experiments for tensile tests with hydrogen charging and the corresponding numerical simulations are presented next.

\section{Numerical Example}\label{example}
\begin{figure}
\begin{center}
\includegraphics[scale=0.25]{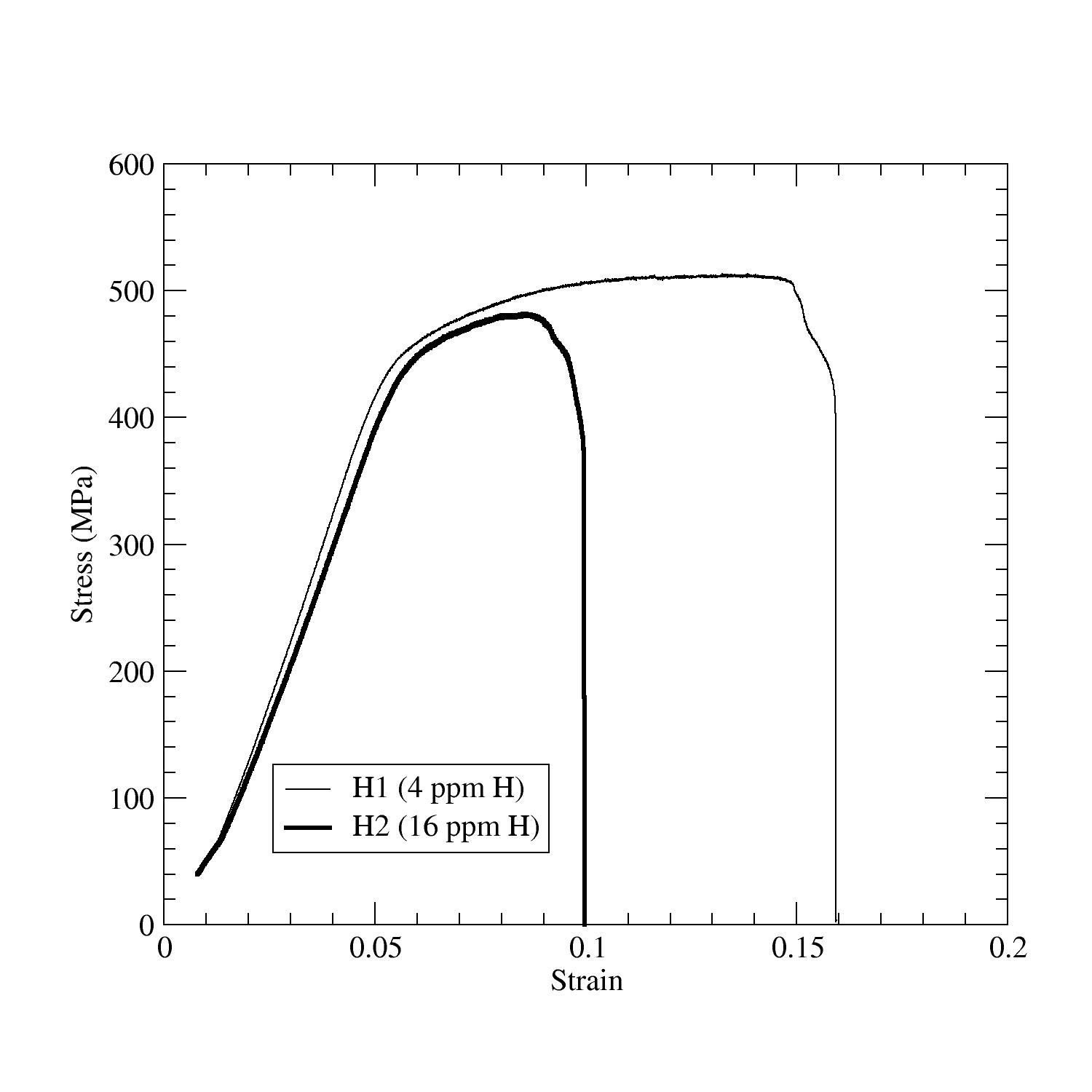}
\end{center}
\caption{Experimental tensile stress-strain response for round bar specimen with two different exposure levels of hydrogen. The hydrogen charging levels are attained for each experiment through electrochemical precharging.}
\label{fig-expts}
\end{figure}

In this section, experiments to study the effect of hydrogen are discussed, followed by numerical simulation of these experiments. Round bar tensile tests are performed on X65 grade linepipe steel with {\it in-air} mechanical properties of $\sigma_o (initial~yield)=420MPa$, Ultimate Tensile Strength (UTS)$ = 520MPa$ and tensile elongation $18\%$. The tensile sample is characterized by a gage length of $2.54cm (1 inch)$ and cross-section diameter $0.3175cm (1/8 inch)$, with an applied tensile strain rate of $10^{-5} /sec$ until failure. For studying the influence of hydrogen on the tensile response, the specimens are electrochemically precharged using $0.5M H_2SO_4, 5mg/l ~ As_2O_3$ for $4 ~ hours$. A current density of $5mA/cm^2$ is used to charge $4 ppm$ (equivalent to $31.5 mol/m^3$) of hydrogen in the sample, as measured through a gas chromatograph(GC). This is denoted as {\it H1} charging condition. A current density of $50mA/cm^2$ is used to charge $16 ppm$ (equivalent to $126 mol/m^3$) of hydrogen in the sample, denoted as {\it H2} charging condition. The tensile test is conducted on the charged specimens in hydrogen gas at $1 atm$ pressure to minimize hydrogen egress during testing. The samples are tested to failure and the tensile response for {\it H1} and {\it H2} conditions is shown in figure \ref{fig-expts}. As inferred from the tensile response, the total elongation to failure ($e_f$) decreased from $18\%$ in air to $15.5\%$ for {\it H1} charging and to $9.5\%$ for {\it H2} charging condition.

\begin{figure}
\begin{center}
\includegraphics[scale=0.25]{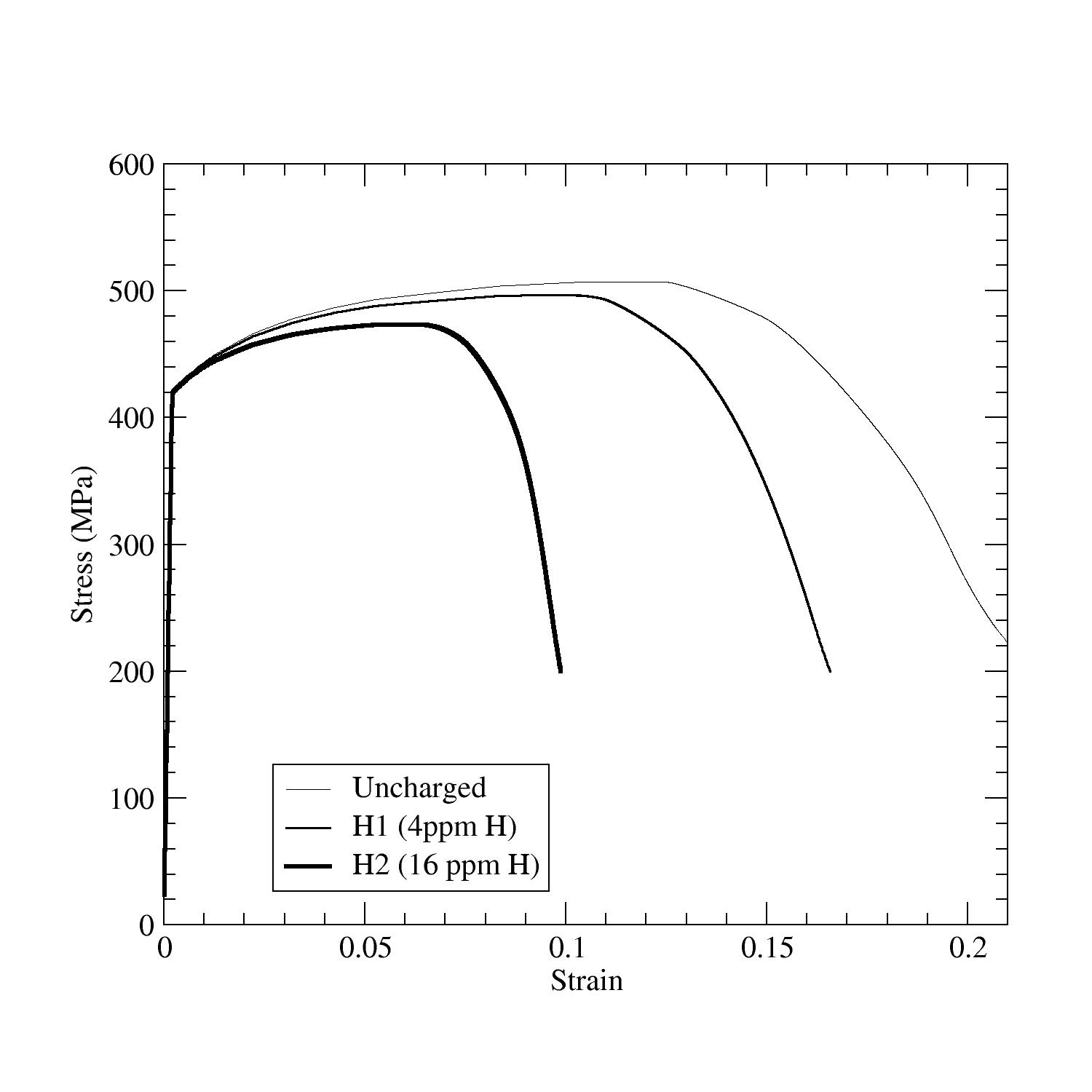}
\end{center}
\caption{Numerical tensile stress-strain response for round bar specimen at various hydrogen exposure levels. The damage model developed in this work is capable of estimating hydrogen induced damage to match with experiments.}
\label{fig-simulation}
\end{figure}

The above experiments for hydrogen induced damage are simulated using the continuum damage model developed in this work. The basic calibration values for {\it in-air} material properties are as mentioned in the previous section. The damage parameter for hydrogen induced softening is chosen as $C_c=1500 mol/m^3$ and the parameter quantifying the hydrogen induced void nucleation is chosen as $B= 3 \times 10^{-3} m^3/mol$. The values of these parameters are identified based on numerical experiments over the parameter space. The round bar tensile specimen under uniaxial tensile loading is simulated with deformation coupled with hydrogen diffusion in the specimen. The resulting engineering stress-strain curves are plotted in figure \ref{fig-simulation}. The NVC damage model is able to estimate the loss of total elongation $e_f$ due to the presence of hydrogen for both the charging conditions.

\section{Discussion}\label{discussion}
The continuum damage model developed in this work adopts the NVC theory proposed by Neeraj and coworkers as the fundamental mechanism for hydrogen embrittlement in ferritic steels. The NVC theory postulates that the hydrogen enhances localized plasticity and stabilizes vacancy agglomeration to nano-voids. These key elements are incorporated in the proposed damage model for hydrogen embrittlement. The damage model is built on Gurson model as a platform. 

The localized plasticity occurs through hydrogen induced localization of dislocation plasticity at the meso-scale in the material. This is translated as contribution of hydrogen leading to softening in the material. The consequent effect is incorporated in the mathematical formulation of the hardening law. Various forms of modified hardening laws have been proposed in the literature for hydrogen induced softening and hardening. While hydrogen has not been observed to enhance hardening in the steels under study, the proposed law solely focuses on enhanced softening. Additional attention is given to the functional form of the hardening law to respect the extreme conditions of no hydrogen and very high local hydrogen concentrations. The law reduces to normal hardening in the absence of hydrogen and translates to perfectly plastic material law at a critical concentration of (local) hydrogen. These effects are incorporated within the hardening law using a single new parameter, denoted as critical hydrogen concentration ($C_c$). 

The vacancy stabilization has been described by NVC theory to be strongly influenced by dislocation plasticity in the presence of hydrogen. The contribution is directly translated to accelerate void nucleation in the material. The consequent effect in the mathematical formulation appeared in the void nucleation formulation as an additional term dependent on both the hydrogen concentration and plastic strain. The enhanced void nucleation is mathematically incorporated with a new parameter denoted as {\it $H-V$ complex coefficient} ($B$). The effect of hydrogen on void formation is quantified with this parameter. As the $H-V$ complex coefficient reaches zero, the void nucleation is dictated solely by carbides splitting and debonding. Non-zero values of the parameter contribute to $H-V$ complex driven void nucleation. The sensitivity of this parameter to the tensile response of round bar specimen is also conducted in this work. 

While there are multiple efforts in the literature to model hydrogen embrittlement, they differ from the proposed model both in terms of the mathematical formulation and fundamental failure mechanism. The damage model developed in this work is oriented towards a new hydrogen influenced constitutive relationship. In contrast to other efforts in the literature, the new constitutive relationship is capable of simulating the continuous damage response of material from {\it no hydrogen} to critical concentrations of hydrogen, within the same framework. In addition, the Gurson based model is a combined damage and yield function. Hence it incorporated both plasticity and damage into the material response. This gave the capability to simulate the response of hydrogen induced damage besides plastic deformation.

As with any continuum damage model, the proposed model suffers from finite element mesh sensitivity. Since the damage occurs at local material points that are not informed of neighboring points, the overall response is localized resulting in mesh sensitivity. An alternative to this method would involve developing a nonlocal damage model that incorporates the influence of damage within a radius of characteristic length at each material point. Since nonlocal models counteract the desired plasticity localization due to hydrogen, the current formulation included a similar length scale parameter associated with the physical defect induced at the specimen center line. While the radius of specimen centerline is reduced by $0.5\%R$ to induce a natural defect, the size of the element along this defect becomes a characteristic material length scale for the model. The model parameters are calibrated with a specific characteristic material length scale and hence would differ with a finer or coarser mesh. The sensitivity of model parameters with the characteristic material length scale is not studied in this work and is identified as part of future work. 

The continuum damage model developed in this work is coupled with hydrogen diffusion model proposed by Sofronis and coworkers. The hydrogen diffusion model incorporated the effect of trap mediated diffusion and hydrostatic stress gradients in the material. The dissolved hydrogen in-turn contributed to the deformation through local dilatation of lattice. This enabled to simulate the effect of the coupled hydrogen transport and hydrogen induced damage on the mechanical behavior of material. While the present work incorporates a single type of trap, the development can be easily extended to multiple traps. Finally, the capability of the coupled damage model is demonstrated by simulating tensile response of commercial linepipe steel under the influence of hydrogen. The numerical simulations closely matched the experimental behavior under two different hydrogen loading conditions.

\section{Summary}\label{summary}
A continuum damage model is developed in this work for simulating hydrogen embrittlement in steels. The model adopts Nano-Void Coalescence theory introduced by Neeraj and coworkers. The damage model encompasses hydrogen enhanced dislocation plasticity and hydrogen stabilized vacancy agglomeration within the frame work of ductile void coalescence theory. The formulation is built on the Gurson ductile constitutive response and is two-way coupled with hydrogen diffusion in the specimen that is influenced by the presence of microstructural traps. The coupled model is implemented within ABAQUS User Material (UMAT) framework wherein stress update, tangent modulus and diffusion calculations are performed. Sensitivity analyses are performed for the new parameters introduced to capture the influence of hydrogen. Finally experimental tensile tests are simulated for round bar specimens under various hydrogen loading conditions. Simulated stress-strain curves are in good agreement with experimental results, thus establishing the capability of the damage model.

\section*{Acknowledgements}
The authors acknowledge the support of ExxonMobil Technology and Engineering to conduct this research and their permission to publish this work.

\bibliography{mybibfile}

\end{document}